\documentclass[12pt]{oxarticle}
\pdfoutput=1

\usepackage{graphicx, float, array, xspace, amscd, amsmath, amsthm, amssymb, latexsym, relsize, bbm}
\usepackage[letterpaper, left=1.7in, right=0.3in, top=0.8in, bottom=1.2in]{geometry}
\usepackage[bbgreekl]{mathbbol}
\usepackage{amsfonts, tikz-cd}
\usepackage[T1]{fontenc}
\usepackage{tikz}
\usetikzlibrary{backgrounds}
\usepackage[framemethod=tikz]{mdframed}
\AtBeginEnvironment{mdframed}{%
\tikzset{every picture/.style={}}%
}
\mdfsetup{roundcorner=.5ex}
{\begin{mdframed}[backgroundcolor=white]}%
{\end{mdframed}}

\DeclareSymbolFontAlphabet{\mathbb}{AMSb}
\DeclareSymbolFontAlphabet{\mathbbl}{bbold}

\usepackage[sort&compress, comma, square, numbers]{natbib}
\usepackage[all,cmtip]{xy}
\usepackage{color}
\definecolor{MyDarkBlue}{rgb}{0.15,0.25,0.45}
\usepackage[linktocpage=true]{hyperref}
\hypersetup{colorlinks=true, citecolor=blue, linkcolor=blue, urlcolor=blue, pdfauthor={}, pdftitle={},breaklinks=true}

\newcommand{\w}{{\,\wedge\,}}
\newcommand{\wt}{\widetilde}

\def\rep#1{{{\boldsymbol{#1}}}}
\def\brep#1{{{\overline{\boldsymbol{#1}}}}}

\def\CS{{\text{CS}}}

\renewcommand{\a}{\alpha}

\newcommand{\D}{\Delta}

\renewcommand{\l}{\lambda}
\newcommand{\m}{\mu}
\newcommand{\n}{\nu}
\newcommand{\x}{\xi}

\renewcommand{\r}{\rho}
\newcommand{\s}{\sigma}
\renewcommand{\t}{\tau}

\DeclareFontFamily{OT1}{pzc}{}
\DeclareFontShape{OT1}{pzc}{m}{it}{<-> s * [1.200] pzcmi7t}{}
\DeclareMathAlphabet{\mathpzc}{OT1}{pzc}{m}{it}

\newcommand{\cA}{\mathcal{A}}

\newcommand{\cF}{\mathcal{F}}

\newcommand{\cR}{\mathcal{R}}

\newcommand{\cT}{\mathcal{T}}

\newcommand{\ccZ}{\mathpzc Z}

\DeclareFontFamily{U}{bbold}{}
\DeclareFontShape{U}{bbold}{m}{n}
 {  <-5.5> s*[1.05] bbold5
    <5.5-6.5> s*[1.05] bbold6
    <6.5-7.5> s*[1.05] bbold7
    <7.5-8.5> s*[1.05] bbold8
    <8.5-9.5> s*[1.05] bbold9
    <9.5-11.5> s*[1.05] bbold10
    <11.5-16> s*[1.05] bbold12
    <16-> s*[1.05] bbold17
 }{}

\renewcommand{\Im}{\mathbbl{m}}

\font\eightrm=cmr8 at 8pt
\font\csc=cmcsc10

\newcommand{\beq}{\begin{equation}}
\newcommand{\eeq}{\end{equation}}
\newcommand{\beqnn}{\begin{equation*}}
\newcommand{\eeqnn}{\end{equation*}}
\newcommand{\bea}{\begin{eqnarray}}
\newcommand{\eea}{\end{eqnarray}}
\newcommand{\bean}{\begin{eqnarray*}}
\newcommand{\eean}{\end{eqnarray*}}

\newcommand{\place}[3]{\vbox to0pt{\kern-\parskip\kern-7pt
                             \kern-#2truein\hbox{\kern#1truein #3}
                             \vss}\nointerlineskip}

\DeclareFontFamily{U}{wncy}{}
\DeclareFontShape{U}{wncy}{m}{n}{<->wncyr10}{}
\DeclareSymbolFont{mcy}{U}{wncy}{m}{n}
\DeclareMathSymbol{\sha}{\mathord}{mcy}{"58}

\newcommand{\del}{{\partial}}
\newcommand{\delb}{{\overline{\partial}}}

\newcommand{\nb}{{\overline\n}}
\newcommand{\mb}{{\overline\m}}

\newcommand{\A}{\cA}
\renewcommand{\aa}{\mathfrak{a}}

\newcommand{\dd}{{\text{d}}}

\newcommand{\K}{K\"ahler\xspace}

\renewcommand{\Im}{\text{Im~}}

\def\ker{{\rm ker ~}}

\newcommand{\tr}{\text{Tr}\hskip2pt}

\newcommand{\ap}{{\a^{\backprime}\,}}

\renewcommand{\=}{\;=\;}

\makeatletter
\g@addto@macro\bfseries{\boldmath}
\makeatother

\newcommand{\citeE}{\cite{McOrist:2021dnd}\xspace}

\newcommand{\citeSG}{\cite{McOrist:2019mxh}\xspace}

\newcommand{\citeUG}{\cite{Candelas:2018lib, McOrist:2019mxh}\xspace}

\renewcommand{\baselinestretch}{1.1}
\numberwithin{equation}{section}
\proofmodefalse
\begin{document}
\pagestyle{empty}      
\ifproofmode\underline{\underline{\Large Working notes. Not for circulation.}}\else{}\fi

\begin{center}
\null\vskip0.2in
{\Huge  The decoupling of moduli about the standard embedding\\[0.5in]}
{\csc   Beatrice Chisamanga$^{\dagger\,1}$, Jock McOrist$^{\sharp\,2}$, Sebastien Picard$^{\$\,3}$,  and Eirik Eik Svanes$^{\dagger\,4}$\\[0.5in]}

{\it 

$^\dagger$ Department of Mathematics and Physics \\
Faculty of Science and Technology\\
University of Stavanger\\
N-4036, Stavanger, Norway\\[3ex]

$^\sharp$Department of Mathematics\hphantom{$^2$}\\
School of Science and Technology\\
University of New England\\
Armidale, 2351, Australia\\[3ex]

$^\$$ Department of Mathematics\\
University of British Columbia\\
 1984 Mathematics Road\\
  Vancouver, BC, Canada\\
  
}

%
\footnotetext[1]{{\tt b.chisamanga@stud.uis.no }}
\footnotetext[2]{{\tt jmcorist@une.edu.au}}
\footnotetext[3]{{\tt spicard@math.ubc.ca}}
\footnotetext[4]{{\tt eirik.e.svanes@uis.no}}
\vspace{1cm}
{\bf Abstract\\[-8pt]}
\end{center}

We study the cohomology of an elliptic differential complex arising from the infinitesimal moduli of heterotic string theory in the supergravity approximation. We compute these cohomology groups at the standard embedding, and show that they decompose into a direct sum of cohomologies. While this is often assumed in the literature, it had not been explicitly demonstrated. Given a stable gauge bundle over a complex threefold with trivial canonical bundle and no holomorphic vector fields, we also show that the Euler characteristic of this differential complex is zero. This points towards a perfect obstruction theory for the heterotic moduli problem, at least for the most physically relevant compactifications.

\vskip150pt

\newgeometry{left=1.5in, right=0.5in, top=0.75in, bottom=0.8in}
%
%
\restoregeometry
\setcounter{page}{1}
\pagestyle{plain}
\renewcommand{\baselinestretch}{1.3}
\null\vskip-10pt

\section{Introduction}
String compactification is a process which embeds a four-dimensional spacetime into a ten-dimensional string theory. The string theory studied in 1986 by \cite{Candelas:1985en} is the heterotic string, whose spacetime has an ansatz of the form 
\begin{equation}
 M_{10-n} \times X_n^{compact}~,
\end{equation} 
where $M_{10-n}$ is non--compact and the vector bundle is identified with the tangent bundle -- this setup is known as the standard embedding. More general heterotic string compactifications, of interest due to their ability to embed chiral gauge theories close to the minimally supersymmetric standard model,  for example \cite{Bouchard:2005ag}, and their tractability in computing quantum corrections which may be  phrased in terms of geometric conditions on $X_n^{compact}$. The most physically relevant case is $n=6$  described in \cite{Hull:1986kz,Strominger:1986uh}. As the string compactification at a fundamental level is described by a two-dimensional conformal field theory, it is possible to replace the geometry of $X$ by an abstract CFT with an appropriate collection of symmetries such as a Gepner model. Nonetheless,  theories defined by Calabi-Yau manifolds with stable vector bundles are still the  most widely studied due to their well established mathematics and being amenable to both worldsheet and supergravity methods. 

We focus here on heterotic compactifications that realise four-dimensional Minkowski spacetime with $N=1$ supersymmetry, meaning that $X$ is a compact complex 3-fold with vanishing first Chern class. There is a vector bundle $V$ that is holomorphic which admits a connection that is hermitian Yang-Mills. The Green-Schwarz anomaly relates the hermitian form on $X$ to topological constraints on $V$.  Over the years many authors have established solutions of the 10-dimensional heterotic supergravity, and realized a 4-dimensional Minkowski space via the standard embedding and other ways. These geometries define a conformal field theory with $(0,2)$ supersymmetry, and little is known about these theories at a fundamental level. There is a limit in which this enhances to $(2,2)$ supersymmetry, with many more calculation tools available, and this limit is known as the standard embedding. It is in this limit that mirror symmetry and special geometry exists with which the complete couplings of the theory can be determined. A simple problem going beyond this limit is to study deformations of the standard embedding that do not necessarily preserve $(2,2)$ supersymmetry. Geometrically, this amounts to studying deformations of the Calabi-Yau manifold together with deformations of the tangent bundle that are not connected to the underlying Calabi-Yau such that the spacetime supersymmetry conditions are satisfied. 

From the point of view of spacetime, the deformations are coupled via the supersymmetry variations and the Bianchi identity. When written in terms of spinor bilinears, and related to the differential geometric structures of the internal manifold $X$, some of these can also be written as the variation of a functional called the superpotential \cite{LopesCardoso:2003dvb,Gurrieri:2004dt,delaOssa:2015maa,McOrist:2016cfl}, so we call them F-terms in analogy with $N=1$ $d=4$ supersymmetry. These were extended in \cite{Ashmore:2018ybe} to a study of finite deformations. The remaining supersymmetry variations we would like to refer to as D-terms, continuing the analogy, and evidence for this was presented in \cite{McOrist:2021dnd}, and further studied in \cite{MPS24} in which using the string theory moduli space metric, calculated in \cite{Candelas:2016usb,McOrist:2019mxh} by a dimensional reduction to first order in $\ap$, it was demonstrated that the F-terms correspond to the kernel of a certain $\overline D$-operator and the D-terms lie in the kernel of its adjoint ${\overline D}^\dag$ with respect to the metric on moduli. In \cite{Ashmore:2019rkx,Garcia-Fernandez:2020awc} the D-terms were presented in terms of a moment map construction, all of this consistent with the usual $N=1$ $d=4$ supersymmetry lore. All this being said, and unlike the case of the moduli space of Calabi-Yau manifolds (e.g. see the classic work \cite{Candelas:1990pi}), it is certainly not obvious the deformations decouple into a direct sum of cohomologies related to complex structure of the manifold, the hermitian structure, and endomorphisms of the tangent bundle. A toy example which we return to later is that, as noted by Atiyah \cite{Atiyah:1955}, as the bundle is holomorphic it requires a field strength $F^{(0,2)}  = 0$ with respect to the complex structure of $X$. For a generic deformation of complex structure, one might worry we generate some $F^{(0,2)}$ that violates this condition; indeed, this is the case if such a deformation is not $\delb_{\mathcal A}$--exact, in which case the corresponding complex structure deformation of the manifold $X$ is not a deformation or parameter of the heterotic theory. See \cite{Anderson:2010mh,Anderson:2011ty,Anderson:2011cza} for applications of this mechanism to the heterotic moduli problem.

From the point of view of the worldsheet theory, when studied via a gauged linear sigma model the deformations of the bundle and the traditional deformations of the CY appear on the same footing at least when computing worldsheet instanton corrections to Yukawa couplings \cite{McOrist:2007kp,McOrist:2008ji}. The same conclusion holds when the semi-classical non--linear sigma model is studied \cite{Melnikov:2011ez}. Thence, it is not obvious what the action of mirror symmetry is on the parameter space. That being said, this question has been answered to some extent for CY manifolds that are complex intersections in toric varieties satisfying a certain combinatorial condition \cite{Kreuzer:2010ph,Melnikov:2010sa}, but the connection to the spacetime description of the moduli space, via the Hull-Strominger system, is an open question. What happens when this condition is not satisfied is also an open question, even for deformations of the standard embedding.

The goal of this paper is to demonstrate that the parameter space of heterotic theories about the standard embedding decomposes into a sum of cohomologies utilising results derived in \citeUG and later in \citeE.  We also draw inspiration from Atiyah's work \cite{Atiyah:1955}, and later applications in physics and mathematics \cite{Anderson:2010mh, Anderson:2011ty, Anderson:2011cza, delaOssa:2014cia, Anderson:2014xha, Garcia-Fernandez:2015hja, Garcia-Fernandez:2020awc, Silva:2024fvl, Kupka:2024rvl}, where the simultaneous deformations of the geometry and the bundle were considered at the level of $\ap$-correct supergravity.

More specifically, we are interested in the total moduli that result from the simultaneous deformations of the metric, complex structure, and gauge connection. The infinitesimal moduli are computed by a cohomology $H^{0,1}_{\bar D}(Q)$ \citeE, which can be computed using homological algebra techniques and long exact sequences. We show that these sequences split at the standard embedding, leading to an infinitesimal spectrum
\begin{equation}
H_{\bar D}^{0,1}(Q) ~\cong~ H^{1,1}(X) \oplus H^{2,1}(X) \oplus H^{0,1}({\rm End}_0(V))~.
\end{equation}
Note, there is a  substantial swath of literature that implicitly use this result, without having actually verified it is true. We do that here. 

Finally, given physically motivated assumptions about the holomorphic bundles involved, i.e. a stable gauge bundle over a complex threefold with trivial canonical bundle and no holomorphic vector fields, we also show that the associated heterotic moduli complex has vanishing Euler characteristic. This is a simplified version of a more general proof of vanishing index for the moduli problem of heterotic $SU(3)$ solutions found in \cite{deLazari:2024zkg}. This points towards a perfect obstruction theory for heterotic moduli, at least for the most physically relevant compactifications, which may also have implications for the heterotic string theory moduli problem, and for understanding geometric invariants in this setting. 

\section{An elliptic complex}
We begin by showing that the differential complex governing the six-dimensional heterotic moduli problem is elliptic, and so the corresponding cohomologies are finite-dimensional. Let $(X,\omega)$ be a complex hermitian manifold. Let $V \rightarrow X$ be a holomorphic vector bundle with connection and denote its curvature 2-form by $F$. Let $Q = T^{*(1,0)}X \oplus {\rm End}_0(V) \oplus T^{1,0}X$. We equip the smooth complex vector bundle $Q \rightarrow X$ with a differential operator 
\beq \label{Dbar-defn}
\bar{D}: \Omega^{0,p}(Q) \rightarrow \Omega^{0,p+1}(Q), \quad \bar{D} = \begin{bmatrix}
\bar{\partial} & \alpha' \cF^* & \cT + \alpha' \cR \cdot \nabla \\
0 & \bar{\partial}_A & \cF \\
0 & 0 & \bar{\partial}
\end{bmatrix}.
\eeq
The definitions are as follows:
\begin{align*}
\mathcal{F}(\Delta) &= F_{\mu \bar{\nu}} \dd x^{\bar{\nu}} \wedge \Delta^\mu \\
\mathcal{F}^*(\mathfrak{a}) &= {\rm Tr} \, F_{\mu \bar{\nu}} \dd x^\mu \otimes \dd x^{\bar{\nu}} \wedge \mathfrak{a} \\
\mathcal{T}(\Delta) &= H_{\rho \bar{\nu} \mu} \dd x^\rho \otimes \dd x^{\bar{\nu}} \wedge \Delta^\mu \\
\mathcal{R} \cdot \nabla (\Delta) &= -\frac{1}{p!}  R_{\rho \bar{\mu}}{}^\sigma{}_\lambda \hat{\nabla}_\sigma \Delta^\lambda{}_{\bar{\kappa}_1 \cdots \bar{\kappa}_p} \, \dd x^\rho \otimes \dd x^{\bar{\mu} \bar{\kappa}_1 \cdots \bar{\kappa}_p}\:,
\end{align*}
where $\Delta \in \Omega^{0,p}(T^{1,0}X)$, $\mathfrak{a} \in \Omega^{0,p}({\rm End}_0(V)$, $H = i (\partial - \bar{\partial})\omega$ and $\hat{\nabla}$ is the Bismut connection when acting on holomorphic ``free" indices, and the Chern connection when acting on form indices. The mathematical origins of this operator can be found in \cite{bismut1989,gualtieri2014} and its relevance in heterotic string theory was discovered in \cite{Anderson:2014xha,delaOssa:2014cia}; see \cite{McOrist:2021dnd} for the setup without spurious modes. The significance of the operator $\bar{D}$ is that its nilpotency is related to the heterotic Bianchi identity.

The calculation of \cite{MPS24} shows that the equation $\bar{D}^2 = 0$ is equivalent to
\beq \label{bianchi-chern}
i \partial \bar{\partial} \omega = \frac{\alpha'}{2} {\rm Tr} \, F \wedge F - \frac{\alpha'}{2} {\rm Tr} \, R \wedge R\:,
\eeq
where $R$ is the Chern curvature of the hermitian metric $\omega$. We note that this is not the usual Bianchi identity from string theory since there $R$ is computed with respect to the Hull connection \cite{Hull:1986kz}, however \eqref{bianchi-chern} agrees with the physical system up to $O(\alpha'^2)$ in the regime where $H=O(\alpha')$, and so \eqref{bianchi-chern} is a good mathematical approximation to the physical system at order $O(\alpha'^2)$ in the action when $H=O(\alpha')$. The advantage of \eqref{bianchi-chern} is that the exact equality $\bar{D}^2=0$ allows us to use methods from homological algebra to compute the cohomology of the complex defined by $\bar{D}$.

In summary, we have interpreted equation \eqref{bianchi-chern} as a sort of holomorphic structure $(Q,\bar{D})$ on the smooth bundle $Q$. However, we note that
\[
\bar{D} (f q) \neq (\bar{\partial} f)q + f \bar{D} q, \quad f \in C^\infty(X), \ q \in \Gamma(Q),
\]
due to the $\alpha' \mathcal{R} \nabla$ off-diagonal corrections. Thus $(Q,\bar{D})$ is not a holomorphic structure in the traditional sense and the discrepancy occurs at order $\alpha'$.

Provided \eqref{bianchi-chern} holds, we have $\bar{D}^2=0$ and the differential operator $\bar{D}$ defined in \eqref{Dbar-defn} defines a differential complex
\beq \label{elliptic-complex}
0 \longrightarrow \Gamma(Q) \overset{\bar{D}}{\longrightarrow} \Omega^{0,1}(Q) \overset{\bar{D}}{\longrightarrow} \Omega^{0,2}(Q) \rightarrow \cdots\:.
\eeq
The current paper initiates the study of the cohomology of this complex. In this section, we will show that this complex is elliptic. For this, we must show that for all $p \in X$, then for all $\xi \in T^*_p X \backslash \{0 \}$ then the symbols
\beq
\sigma(\bar{D},\xi)|_p: \Omega^{0,k}(Q)|_p \rightarrow \Omega^{0,k+1}(Q)|_p
\eeq
have the property
\beq
{\rm Ker} \, \sigma(\bar{D},\xi)|_p = {\rm Im} \, \sigma(\bar{D},\xi)|_p\:.
\eeq
The vector $\xi$ is real, which we write in complex coordinates as $\xi = \xi_\mu d x^\mu + \xi_{\bar{\mu}} d x^{\bar{\mu}}$ with $\xi_{\bar{\mu}} = \overline{\xi_\mu}$. The symbol $\sigma(\bar{D},\xi)$ is given by
\beq
\sigma(\bar{D},\xi) = \begin{bmatrix} \sigma(\bar{\partial},\xi) & 0 & A(\xi)\\
0 & \sigma(\bar{\partial},\xi) & 0\\
0 & 0 & \sigma(\bar{\partial},\xi)
\end{bmatrix},
\eeq
with
\beq
\sigma(\bar{\partial},\xi) \ccZ = \xi_{\bar{\mu}} \frac{\ccZ_{\alpha \bar{K}}}{k!} \dd x^\alpha \otimes \dd x^{\bar{\mu}} \wedge \dd x^{\bar{K}}, \quad A(\xi) \D = - R_{\alpha \bar{\mu}}{}^\sigma{}_\lambda \xi_\sigma \frac{\D^\lambda{}_{\bar{K}}}{k!} \dd x^\alpha \otimes \dd x^{\bar{\mu}} \wedge \dd x^{\bar{K}}.
\eeq
Since $\sigma(\bar{D},\xi) \sigma(\bar{D},\xi) = 0$, we have ${\rm Im} \, \sigma(\bar{D},\xi)|_p \subseteq {\rm Ker} \, \sigma(\bar{D},\xi)|_p$. To show ellipticity we take $q \in \Omega^{0,k}(Q)$ with
\beq
q = \begin{bmatrix} \ccZ \\ \aa \\ \D \end{bmatrix} \in \ker \sigma(\bar{D},\xi)
\eeq
and show $q \in {\rm Im} \, \sigma(\bar{D},\xi)|_p$. Since $\sigma(\bar{\partial},\xi)$ is elliptic, then
\bea
\D &=& \sigma(\bar{\partial},\xi) \hat{\D}, \quad \hat{\D} \in \Omega^{0,k-1}(Q) \\
\aa &=& \sigma(\bar{\partial},\xi) \hat{\aa}, \quad \hat{\aa} \in \Omega^{0,k-1}(Q).
\eea
The constraint $\sigma(\bar{D},\xi)q = 0$ implies
\beq
\sigma(\bar{\partial},\xi) \ccZ + A(\xi) \D = 0.
\eeq
This is
\beq
0 = \sigma(\bar{\partial},\xi) \ccZ + A(\xi) \sigma(\bar{\partial},\xi) \hat{\D} = \sigma(\bar{\partial},\xi) (\ccZ - A(\xi) \hat{\D})
\eeq
since $A(\xi) \sigma(\bar{\partial},\xi) = - \sigma(\bar{\partial},\xi) A(\xi)$. Ellipticity of $\sigma(\bar{\partial},\xi)$ implies
\beq
\ccZ - A(\xi) \hat{\D} = \sigma \hat{\ccZ}.
\eeq
Altogether,
\beq
 \begin{bmatrix} \sigma(\bar{\partial},\xi) & 0 & A(\xi)\\
0 & \sigma(\bar{\partial},\xi) & 0\\
0 & 0 & \sigma(\bar{\partial},\xi)
\end{bmatrix} \begin{bmatrix} \hat{\ccZ} \\ \hat{\aa} \\ \hat{\D} \end{bmatrix} = \begin{bmatrix} \ccZ \\ \aa \\ \D \end{bmatrix}
\eeq
as desired.

From the elliptic complex \eqref{elliptic-complex}, we can define the cohomology groups
\beq
H^{0,q}_{\bar{D}}(Q) = \frac{{\rm Ker} \, ( \bar{D}: \Omega^{0,q}(Q) \rightarrow \Omega^{0,q+1}(Q))} {{\rm Im} \, ( \bar{D}: \Omega^{0,q-1}(Q) \rightarrow \Omega^{q,0}(Q))}.
\eeq
On the space $\Omega^{0,q}(Q)$, we may introduce the $L^2$ inner product:
\bea
g(q_1,q_2) &=& \int_X    \frac{1}{4}  g^{\mu \bar{\nu}} (\ccZ_1)_\mu \star \overline{(\ccZ_2)_\nu} - \frac{\ap}{4} {\rm Tr} \, (\aa_1 \star \aa_2^\dagger) + g_{\mu \bar{\nu}}\D_1^\mu \star \overline{\D_2^\nu} \nonumber\\
&&+ \frac{\alpha'}{2}  \int_X R^{\bar{\mu} \rho \sigma \bar{\nu}} \bigg( (-1)^{q+1} \frac{1}{4}  (\ccZ_1)_{\rho \bar{\mu}} \star \overline{(\ccZ_2)}_{\sigma \bar{\nu}} + (-1)^{q} (\D_1)_{\bar{\mu} \bar{\nu}} \overline{(\D_2)}_{\rho \sigma} \bigg)\:,
\eea
where, to correct to first order in $\alpha'$ the curvature $R$ in the last line is the Riemann curvature. When $q=0$, the inner product does not include the last line. These $\alpha'$-corrections on the last line are derived from the moduli space metric on the heterotic moduli space and are due to \cite{Candelas:2016usb, Candelas:2018lib, McOrist:2021dnd}. For $\alpha'$ small enough, the expression defines a positive-definite $L^2$ inner product.

This inner product allows us to define the adjoint operator
\beq
\bar{D}^\dagger : \Omega^{0,q}(Q) \rightarrow \Omega^{0,q-1}(Q)\:,
\eeq
and form the Laplacian
\beq
\Box = \bar{D}^\dagger \bar{D} + \bar{D} \bar{D}^\dagger\:.
\eeq
Since the complex is elliptic, we obtain finite dimensional spaces of harmonic forms.
\beq
\mathcal{H}^q(Q) = {\rm Ker} \, (\Box: \Omega^{0,q}(Q) \rightarrow \Omega^{0,q+1}(Q))\:.
\eeq
The significance of these spaces is that the analysis in \cite{McOrist:2021dnd} shows that $\mathcal{H}^1(Q)$ parametrizes the heterotic moduli. By the Hodge theorem for elliptic complexes (see e.g. \cite{wells1980}), we conclude
\beq
\dim \mathcal{H}^1(Q) = \dim H_{\bar{D}}^{0,1}(Q)\:,
\eeq
and this space is finite dimensional by ellipticity of the complex. This correspondence allows us to use methods from homological algebra to compute $\dim \mathcal{H}^1(Q)$. We give an application of this method to the standard embedding in the following section.

\section{The standard embedding}
We now compute the cohomology $H_{\bar{D}}^{0,1}(Q)$ for the particular example of the standard embedding. The standard embedding \cite{Candelas:1985en}, was the first consistent string compactification. It was constructed with a view to building a theory that incorporated the standard model and gravity. It is defined by taking $X$ to be a CY manifold and embedding the spin connection, an $SU(3)$ connection, in the $E_8\times E_8$ gauge bundle. The $E_8 \times E_8$ is Higgsed to $E_6\times E_8$ and so as the connection takes values in the adjoint representation $\rep{248}$ of $E_8$, it decomposes under $SU(3) \times E_6 \subset E_8$ as
\begin{equation}
    \rep{248} \= (\rep{3}, \rep{27}) \oplus (\brep{3}, \brep{27}) \oplus (\rep{1}, \rep{78}) \oplus (\rep{8},\rep{1})~,
\end{equation}
where the $\rep{78}$ and $\rep{8}$ are the adjoint representation of $E_6$ and $SU(3)$ respectively, and $\rep{27}$ and $\rep{3}$ are their respective fundamental representation with their respective conjugates $\brep{3}$ and $\brep{27}$.  See any of the canonical texts such as \cite{Green:2012pqa} for more details. For the moduli analysis we are only interested in the $SU(3)$ adjoint representation, which are singlets under the external $E_6\times E_8$ gauge symmetry.

Analysing simultaneous deformations of the gauge connection and the complex structure leads to potential obstructions of the complex structure moduli. This is because not all complex structure deformations need preserve the bundle remains holomorphic, as required by supersymmetry. This was first observed by Atiyah \cite{Atiyah:1955}, and applied to the study of heterotic moduli in \cite{Anderson:2010mh, Anderson:2011ty, Anderson:2011cza}. Specifically, the deformations should satisfy the equation
\begin{equation}
\label{eq:AtiyahOb}
\D^\m F_{\m\nb} \dd x^\nb \= \delb_\A \aa~,
\end{equation}
where $F = \dd A + A^2$, $\A  = A^{(0,1)}$,  $\D\in H^{0,1}(T^{(1,0)})$ is the complex structure deformation, $\aa=\delta \A\in\Omega^{0,1}({\rm End}_0(V))$ is the deformation of the gauge connection. This can be rephrased as requiring that the simultaneous deformations $(\aa, \D)$ be in the kernel of 
\begin{equation}
\delb_1=\begin{pmatrix}
\delb _\A & \mathcal{F} \\
0 & \delb 
\end{pmatrix} ~,    \quad \quad \mathcal{F}(\Delta) = F_{\mu \bar{\nu}} \dd x^{\bar{\nu}} \wedge \Delta^\mu
\end{equation}
acting on sections of the complex vector bundle $Q_1 = {\rm End}_0(V) \oplus T^{(1,0)} X$. Here $\mathcal{F}$ is the extension map given by the curvature $F$, often called the Atiyah map. The operator $\bar{\partial}_1$ defines a holomorphic structure $(Q_1,\bar{\partial}_1)$ and forms the differential of the central complex of the Atiyah sequence
\begin{equation}
        0\rightarrow\Omega^{0,q}({\rm End}_0(V)) \xrightarrow{\iota} \Omega_{\delb_1}^{0,q}(Q_1) \xrightarrow{p} \Omega^{0,q}(T^{(1,0)}X)\rightarrow 0~.
\end{equation}
The short exact sequence gives a long exact sequence in cohomology
\begin{equation}
\label{eq:AtiyahLong}
   0 \rightarrow\: H^{0,1}({\rm End}_0(V))\rightarrow H^{0,1}(Q_1)\rightarrow H^{0,1}(T^{(1,0)}X)\xrightarrow{\mathcal{F}} H^{0,2}({\rm End}_0(V)) \rightarrow \cdots~,
\end{equation}
where the sequence begins at level one as the holomorphic tangent bundle of a simply connected K\"ahler Calabi-Yau manifold has no holomorphic sections. The snake lemma gives us that the connecting homomorphism $\mathcal{F}$ is the induced map on cohomology coming from $\mathcal{F}: \Omega^{0,q}(T^{(1,0)}X) \rightarrow \Omega^{0,q+1}({\rm End}_0(V))$ in the $\bar{\partial}_1$ operator given above.

Using the properties of exact sequences, one gets:
\begin{equation}
    H^{0,1}(Q_1) \cong H^{0,1}({\rm End}_0(V)) \oplus \ker\cF~, 
\end{equation}
where $\ker\cF \subseteq H^1(T^{(1,0)}X)$, but does not span this space in general, reflecting the fact that some complex structure moduli are obstructed. 

However, in the case of the standard embedding, the extension map $\mathcal{F}$ is trivial and the long exact sequence splits. To see this, note that the curvature $F$ is in this case given by the Riemann curvature $R$ of the Calabi-Yau. Equation \eqref{eq:AtiyahOb} then takes the form
\begin{equation}
\label{eq:SEAtiyah}
    \dd x^\mb R_{\mb\n}{}^\s{}_\t \w \D^\n \= \delb_\A \aa^\s{}_\t~.
\end{equation}
Let us postulate the following form of $\alpha$
\begin{equation}
\label{eq:alpha}
\aa^\s{}_\t \= \nabla_{\t}\D^\s + \aa_0^\s{}_\t~,    
\end{equation}
where we take $\delb_\A\aa_0^\s{}_\t = 0$, so that $[\aa]\in H^{0,1}({\rm End}_0(V))$ corresponds to a bundle modulus. These are deformations of the gauge bundle which can deform the geometry away from the standard embedding. Using the expression for $\aa^\s{}_\t $, we see that
\beq
\begin{split}
     \delb_\A \aa^\s{}_\t &\=\delb_\A(\nabla_{\s}\D^\t + {\aa_0}^\s{}_\t) \\
    &\=[\delb_\A, \nabla_\t]\D^\s\\
    &\= \dd x^\mb R_{\mb \t}{}^\s{}_\r\wedge \D^\r\= \dd x^\mb R_{\mb \r}{}^\s{}_\t\wedge \D^\r~,\\
\end{split}\notag
\eeq
where we used symmetries of the Riemann curvature on  \K manifold.  This demonstrates that \eqref{eq:AtiyahOb} can always be solved about the standard embedding. Moreover, we have an explicit expression for $\aa$ which we will use below. It follows that the Atiyah extension map ${\cal F}$ is trivial, so that $\ker\cF=H^{0,1}(T^{(1,0)}X)$ and
\begin{equation}
    H^{0,1}(Q_1) \cong H^{0,1}({\rm End}_0(V)) \oplus H^{0,1}(T^{(1,0)}X)~, \notag
\end{equation}
which counts the simultaneous deformations of the complex structure and bundle at the standard embedding.

Solutions of heterotic supergravity are also requred to satisfy the heterotic anomaly cancellation condition
\begin{equation}
    \label{eq:anomaly}
    i(\del -\delb)\omega=\dd B-\frac{\alpha'}{4}\left(\CS(A)-\CS(\nabla)\right)~,\\[2pt]
\end{equation}
where $\omega$ is the hermitian two-form on $X$, and $\nabla$ is a connection on the tangent bundle. This connection is the Levi-Civita connection at the standard embedding, where it is identified with $A$. With $\omega$ being \K, and $\dd B=0$, \eqref{eq:anomaly} is trivially satisfied at this locus. In \citeE it was shown that for simultaneous deformations of the geometry and bundle $(\ccZ,\aa,\D)$, where $\ccZ$ denotes complexified hermitian deformations, to also satisfy anomaly cancellation, they must be in the kernel of the differential \eqref{Dbar-defn} given above, which we can also write as
\begin{equation}\label{ex}
\Bar{D} = \begin{pmatrix}
\delb & \mathcal{H} \\
0 & \delb _1
\end{pmatrix}~.\notag
\end{equation} 
Here the extension map $\cal H$ is determined by the anomaly cancellation. At the standard embedding, the expression \eqref{Dbar-defn} becomes
\begin{equation}
    \bar D = 
\begin{pmatrix}
\delb & \mathcal{F}^* &  \:R \cdot \nabla\\
0 & \delb_\A & \mathcal{F}\\
0 & 0 & \delb 
\end{pmatrix}~,\notag
\end{equation}
acting on $Q=T^{*(1,0)}X \oplus Q_1$. This time $(Q,\bar{D})$ is not a holomorphic structure, however this operator is still the differential of the middle part of a short exact sequence of differential modules
\begin{equation}\label{short2}
0 \rightarrow \Omega^{0,p}(T^{*(1,0)}X) \xrightarrow{\iota} \Omega^{0,p}_{\Bar{D}}(Q) \xrightarrow{p} \Omega^{0,p}(Q_1) \rightarrow 0~.
\end{equation}
As mentioned above, the infinitesimal deformations of the system are counted by $H^{0,1}_{\bar D}(Q)$. The short exact sequence creates a long exact sequence in cohomology
\begin{equation}
    \label{eq:long2}
0\rightarrow H^{0,1}(T^{*(1,0)}X)\rightarrow H^{0,1}(Q)\rightarrow H^{0,1}(Q_1)\xrightarrow{\mathcal{H}} H^{0,2}(T^{*(1,0)}X)\rightarrow\cdots~.   
\end{equation}
Again, the snake lemma identifies the connecting homomorphism $\mathcal{H}$ as induced by $\bar{D}$ and the sequence begins at level one as at the standard embedding $H^0({\rm End}_0(V))=H^0({\rm End}_0(TX))=0$, which also implies $H^0(Q_1)=0$. Using exactness, we can deduce that:
\begin{equation}\label{main}
       H^{0,1}(Q) \cong H^{0,1}(T^{*(1,0)}X) \oplus \ker(\mathcal{H})~,
\end{equation}
where $\ker(\mathcal{H}) \subseteq H^{1,0}(Q_1)$ are the complex structure and bundle deformations satisfying anomaly cancellation. 

We want to show that the extension map $\cal H$ is trivial at the standard embedding. To do so, note again that $F=R$ in this case. For $(\ccZ,\aa,\D)$ to be in the kernel of $\bar D$ implies the following schematic equations
\begin{align}\notag
\delb \ccZ +\mathcal{H}(\D,\aa)=\delb \ccZ + R^* \aa + R \cdot \nabla \D &= 0\\
\delb_\A \aa + R\D &= 0\\
\delb \D &= 0~.    
\end{align}
The last two equations are the conditions to be in the kernel of $\delb_1$, while the first equation written out is
\begin{equation}
   \delb \ccZ +\mathcal{H}(\D,\aa) =\delb \ccZ- \frac{\ap}{2} \tr(F \wedge \aa) + \frac{\ap}{2}R^\s{}_\t \nabla_\s \D^\t \= 0~.
\end{equation}
For $\cal H$ to be trivial, we must show that any pair $(\aa,\D)$ in the kernel of $\delb_1$ satisfies this equation. Consider $\aa$ as in \eqref{eq:alpha}. Taking $R = F$, the above equation can be written as
\beq
\begin{split}
 \delb \ccZ &\=     \frac{\ap}{2}R^\s{}_\t(\nabla_\s \D^\t + {\aa_0}^\t{}_\s) - \frac{\ap}{2}R^\s{}_\t \nabla_\s\D^\t \\
   &\=  \frac{\ap}{2}R^\s{}_\t \: \nabla_\s \D^\t + \frac{\ap}{2}R^\s{}_\t\:{\aa_0}^\t{}_\s - \frac{\ap}{2}R^\s{}_\t \nabla_\s\D^\t \=     \frac{\ap}{2}R^\s_\t\:{\aa_0}^\t{}_\s ~.
\end{split}
\eeq
What this shows is that we need to demonstrate the last term is trivial in cohomology for any $[\aa_0]\in H^{0,1}({\rm End}_0(V))$. We do this now. 

Our strategy is to take the inner product with a generic harmonic form $\Bar{\chi}\in \mathcal{H}^{(1,2)}$. If the inner product is zero then  $R^\s{}_\t{\aa_0}^\t{}_\s$ is $\delb $ exact which follows from the orthogonal decomposition of forms in the Hodge decomposition theorem. Consider 
\begin{equation}\notag
    \langle  R^\s{}_\t{\aa_0}^\t{}_\s, \Bar{\chi}\rangle = \int R ^\s{}_\t\, \aa_0^\t\:{}_\s \wedge \star \chi~,
\end{equation}
where $*\chi$ is a harmonic $(2,1)$-form, which can be written as
\begin{equation}\notag
    \star \chi = \frac{1}{2} \wt\D^\m \:\Omega_{\m\n\l} \dd x^{\n\l}~,
\end{equation}
where $\wt{\D} \in \mathcal{H}^{(0,1)}(T^{(1,0)})$. The inner product becomes
\begin{equation}\notag
\langle  R^\s{}_\t \, {\aa_0}^\t{}_\s, \Bar{\chi}\rangle = \frac{1}{2}\int \tr(R \wedge \aa_0)\:\wedge\:\wt\D^\m\:\wedge\:\Omega_{\m\n\l}\dd x^{\n\l}\:.
\end{equation}
Next, note that the total anti-symmetrisation of four holomorphic indecies vanishes, so that
\begin{equation}\notag
   0 = R_{[\x }\Omega_{\m\n\l]}dz^{\n\l\x} = \frac{1}{4}(3\:R_\x\Omega_{\m\n\l} - R_\m\Omega_{\x\n\l}) \dd x^{\n\l\x}~.
\end{equation}
Using this, we find
\begin{equation}
   \langle  R^\s{}_\t\, {\aa_0}^\t{}_\s, \Bar{\chi}\rangle =   \int tr(R_\m \wedge \aa_0) \wedge \wt\D^\m \wedge \Omega\:.
\end{equation}
But we already know that $R_\m\wt\D^\m = \delb_\A$-exact, as $\wt\D^\m$ corresponds to a complex structure deformation. It follows that the inner product vanishes. We therefore conclude that $\mathcal{H}$ is trivial as a map in cohomology, that is
\begin{equation}
    \ker{\mathcal{H}} = H^{0,1}(Q_1) \cong H^{0,1}(T^{(1,0)}) \oplus H^{0,1}({\rm End}_0(V))~.
\end{equation}
Therefore,
\begin{equation}
    H^{0,1}(Q) \cong H^{0,1}(T^{*(1,0)}X) \oplus H^{0,1}({\rm End}_0(V)) \oplus H^{0,1}(T^{(1,0)}X)~. 
\end{equation}
This proves that in the standard embedding, the total heterotic moduli are given by, \K moduli $H^{1,1}$, complex structure moduli $H^{2,1}$, and bundle moduli $H^{0,1}({\rm End}_0(V))$, as claimed  in the literature.

\section{The index of the complex and obstructions}
As mentioned above, it was shown in \cite{McOrist:2021dnd} that, to first order in $\alpha'$, $H^{(0,1)}_{\bar{D}}(Q)$ computes the infinitesimal spectrum of six-dimensional heterotic solutions. Though not yet demonstrated, it is expected that $H^{(0,2)}_{\bar{D}}(Q)$ will parameterise the obstruction space for the given deformations. If this space has the same dimension as the infinitesimal moduli space the deformation problem is said to have expected or "virtual" dimension zero, meaning the expected moduli space is a set of points. This is interesting not only from the point of view of physics and the string theory moduli problem, but also mathematically in the context of enumerative geometry, invariant theory, and perfect obstruction theories \cite{behrend2009, pandharipande201413}. In this section we demonstrate that, given some physically motivated assumptions, $H^{(0,1)}_{\bar{D}}(Q) \cong H^{(0,2)}_{\bar{D}}(Q)$ and so in the case at hand the space of deformations does have virtual dimension zero. The argument is the "simplified case" version of a more general argument found in \cite{deLazari:2024zkg}, where the physically motivated assumptions are omitted. 

We let $X$ be a compact complex manifold of dimension 3 with nowhere vanishing holomorphic volume form $\Omega$ satisfying
\begin{equation} \label{no-holo}
    H^0(T^{(1,0)} X) = 0\:.
\end{equation}
Let $V \rightarrow X$ be a holomorphic vector bundle with 
\begin{equation} \label{no-holo2}
    H^0({\rm End}_0(V)) = 0\:.
\end{equation}
For example, we may take $V$ to be a stable bundle since these do not admit holomorphic endomorphisms. In this section we will show that in this setup, then
\begin{equation} \label{serre-duality}
\dim H^{0,1}_{\bar{D}}(Q) = \dim H^{0,2}_{\bar{D}}(Q)\:,
\end{equation}
where $H^{0,q}_{\bar{D}}(Q)$ refers to the cohomology of the complex \eqref{elliptic-complex}. By comparing the zeroth and third order cohomology, we will also see that
\begin{equation}
\label{Index}
    \chi(X,\bar{D}) := \sum_{k=0}^3 (-1)^k \dim H^{0,k}_{\bar{D}}(Q)= 0\:,
\end{equation}
and so the Euler characteristic of the complex vanishes. 

{\bf Remark:} We make a few remarks about the assumption \eqref{no-holo}. First, we note that the standard isomorphisms
\begin{equation} \label{hodgenumber}
H^0(T^{(1,0)} X) \cong H^{3,0}(T^{(1,0)} X) \cong H^{0,3}(T^{*(1,0)} X)^*
\end{equation}
hold in the non-K\"ahler setting by the existence of a holomorphic top form $\Omega$ and Serre duality. If $X$ is a K\"ahler Calabi-Yau threefold with $\dim H^1(X,\mathbb{R})=0$, then
\[
\dim H^0(T^{(1,0)} X) = h^{1,3}(X) = h^{3,1}(X) = h^{0,1}(X) = 0\:,
\]
and so assumption \eqref{no-holo} is valid for K\"ahler threefolds. The argument above does not go through for non-K\"ahler manifolds, as it uses the K\"ahler properties $h^{1,3}=h^{3,1}$ and $h^{0,1} \leq \dim H^1(X,\mathbb{R})$.

Nevertheless, we note that \eqref{no-holo} is a natural assumption in the context of non-K\"ahler conifold transitions. If one considers the web of all complex manifolds connected to K\"ahler Calabi-Yau threefolds by conifold transition, then some of these objects may not support any K\"ahler metric, however the condition \eqref{no-holo} still holds through conifold transitions. This is because after a contraction and smoothing $X \rightarrow X_0 \rightsquigarrow X_t$, if infinitely many $X_t$ have $H^0(T^{(1,0)}X_t) \neq 0$ then after normalization one can take a limit of these holomorphic vector fields and obtain a non-zero vector field $W \in H^0(T^{(1,0)} X_{0,{\rm reg}})$ which extends by Hartog's theorem to a holomorphic vector field $W \in H^0(T^{(1,0)}X)$. But since $X$ is a simply connected K\"ahler threefold, such a $W$ does not exist. It follows that for small $t$, the deformed non-K\"ahler object $
X_t$ still satisfies \eqref{no-holo}. For further discussion on conifold transitions, see e.g. \cite{Candelas:1988di,Candelas:1989ug,Candelas:1989jb, Candelas:1990qd,Candelas:1989js,Green:1988bp, Strominger:1995cz, Candelas:2007ac, CPY21,FPS24,FLY12} and references therein. 

We now prove \eqref{serre-duality}. To shorten the notation a bit, let us write $T = T^{(1,0)}X$ and $H^1(Q) = H^{0,1}_{\bar{D}}(Q)$. The long exact sequence \eqref{eq:long2} gives
\[
0 \rightarrow H^1(T^*) \rightarrow H^1(Q) \rightarrow H^1(Q_1) \xrightarrow{\mathcal{H}} H^2(T^*) \rightarrow H^2(Q) \rightarrow H^2(Q_1) \rightarrow 0\:.
\]
since \eqref{hodgenumber} and \eqref{no-holo} imply $H^{0,3}(T^{*(1,0)}X)=0$, and $H^0(Q_1)=0$ holds by \eqref{no-holo}, \eqref{no-holo2}. We conclude
\bea
h^1(Q) &=& h^1(T^*) + \dim \ker \mathcal{H}\:, \label{h1Q} \\
h^2(Q) &=& h^2(Q_1) + h^2(T^*) - \dim \Im \mathcal{H}\:. \label{h2Q}
\eea
where $h^1(Q) = \dim H^1(Q)$. We can apply the same analysis to the long exact sequence \eqref{eq:AtiyahLong} for $Q_1$, which gives
\[
 0 \rightarrow\: H^{1}({\rm End}_0(V))\rightarrow H^{1}(Q_1)\rightarrow H^{1}(T)\xrightarrow{\mathcal{F}} H^{2}({\rm End}_0(V)) \rightarrow H^2(Q_1) \rightarrow H^1(T) \rightarrow 0
\]
since $H^{0,3}({\rm End}_0(V))=0$ by \eqref{no-holo2} and Serre duality, and so we can conclude
\begin{equation} \label{h2Q1}
h^2(Q_1) = h^2(T) + h^2({\rm End}_0(V)) - \dim \Im \mathcal{F}\:.
\end{equation}
Combining \eqref{h2Q} and \eqref{h2Q1} gives
\[
h^2(Q) = h^2(T) + h^2({\rm End}_0(V))+ h^2(T^*) - \dim \Im \mathcal{F} - \dim \Im \mathcal{H}\:.
\]
Using the holomorphic volume form and Serre duality gives
\bea
h^2(T) &=& h^{2,2} = h^{1,1} = h^1(T^*)\:, \nonumber\\
h^2(T^*) &=& h^{1,2} = h^{2,1} = h^1(T)\:, \nonumber\\
h^2({\rm End}_0(V)) &=& h^1({\rm End}_0(V))\:.
\eea
Thus
\[
h^2(Q) = h^1(T^*) + h^1({\rm End}_0(V))+ h^1(T) - \dim \Im \mathcal{F} - \dim \Im \mathcal{H}\:.
\]
We can now substitute
\[
h^1({\rm End}_0(V)) + h^1(T) - \dim \Im \mathcal{F} = h^1({\rm End}_0(V)) + \dim \ker \mathcal{F} = h^1(Q_1) 
\]
to conclude
\[
h^2(Q) = h^1(T^*) + h^1(Q_1) - \dim \Im \mathcal{H}\:,
\]
which agrees with the expression for $h^1(Q)$ given in \eqref{h1Q}. This completes the proof of \eqref{serre-duality}.

It should be noted that the above proof does not rely on the details of the extension map $\mathcal{H}$, so it is also true for the moduli problems of \cite{Anderson:2014xha, delaOssa:2014cia}, where extra spurious modes are included, corresponding to deformations of a connection on the tangent bundle. 

The vanishing of the index, or Euler characteristic \eqref{Index}, then also follows. Indeed, as $H^0(Q_1)=0$, we have 
\[
h^0(Q)=h^0(T^*)=h^3(T)=h^3(Q)\:,
\]
where the middle equality again uses the holomorphic volume form and Serre duality. It follows that the alternating sum in \eqref{Index} vanishes.

\section{Conclusion}\label{conclusion}
We have shown that, at the level of supergravity, correct to first order in $\ap$, the dimension of the parameter space for a point on the standard embedding is 
\beq\label{eq:param}
   h^{1,1}(X)\:+\:h^{2,1}(X)\:+\:h^{0,1}({\rm End}_0(TX))\:.
\eeq
There is the question of extending this infinitesimal analysis to all orders in the deformation theory; namely to rigorously construct coordinates on the moduli space nearby the standard embedding. Progress in this direction in the mathematics literature can be found in \cite{PW24, Garcia_Fernandez_2022}. While it is often assumed that the parameter space of heterotic string theory decomposes into a sum of cohomologies, we have actually demonstrated this here in the particular situation of the standard embedding (the more general situation is much harder and beyond the scope of our current tools). 

This calculation might be interesting in trying to understand the meaning of $(0,2)$ mirror symmetry. As we are considering theories at the standard embedding there is $(2,2)$--mirror symmetry. This is a symmetry of the full string theory and so the dimension of the moduli space of the mirror should at least match that of the original theory. While this is true for the deformations that preserve the $(2,2)$ locus via
$$
h^{1,1}(X) = h^{2,1}(\wt X)~, \qquad {\rm and} \qquad h^{2,1}(X) = h^{1,1}(\wt X)~,
$$
we haven't had a complete expression at the level of supergravity, even for theories that are deformations of the standard embedding, until now. This is of interest for two reasons. First, mathematically speaking it is  certainly not obvious that $h^{0,1}(X, {\rm End}_0(TX))  = h^{0,1}(\wt X, {\rm End}_0(T \wt X))$. So the expression \eqref{eq:param} for $X$ and its mirror $\wt X$ are not necessarily the same. Secondly,  analogous calculations at the level of the worldsheet indicate that the dimension of the moduli space, and its purported mirror, do not agree \cite{Kreuzer:2010ph}. This non-agreement is reflected here.  What do we make of this? 

The moduli we computed in \eqref{eq:param} are massless singlet fields of a supergravity theory correct to first order in $\ap$. It could be the case that the true dimension  of the string theory moduli space differs, modified by worldsheet instantons or $g_s$ corrections. For this reason one might study a sigma model description of the same background, and count the corresponding parameter space. Indeed, this was done in \cite{Kreuzer:2010ph} for CY manifolds that are complete intersections in toric varieties. Unless, $X$ satisfied a certain combinatorial condition (the Newton polytope and its dual both have no facets with interior points) the dimension of the entire $(0,2)$ parameter space for $X$ is not the same as its mirror $\wt X$. In the situation where $X$ does satisfy this combinatorial condition, then the parameter spaces of both $X$ and $\wt X$ are the same and \cite{Melnikov:2011ez} constructed a mirror map on the parameters. It would be interesting, even if this restricted case, to compare the expressions for the parameter space of the GLSM to what is computed in \eqref{eq:param}. They do not necessarily have to agree as, for example the GLSM misses moduli that are  ``non-polynomial'' or ``non-toric" but the moduli it does counts are true at least in the sense they are not lifted by quantum corrections at least in $\ap$. 

It would also be interesting to calculate a similar splitting of cohomologies for more generic heterotic theories. Indeed, the crux of our calculation used the fact that deformations of the gauge connection could be expressed in terms of the moduli $\D$ and $\ccZ$ as per the calculation in \citeSG. We do not have this freedom in more generic situations.  

The computation above also suggests that the Hull-Strominger system has a zero-dimensional virtual moduli space, i.e. the expected moduli space is a set of points. This is the scenario where techniques of enumerative geometry, or counting topological invariants, naturally applies. In this setting, it is tempting to speculate whether analogs of Donaldson-Thomas invariants \cite{thomas1997gauge, donaldson1998gauge} may be defined for heterotic geometries. Note that in \cite{deLazari:2024zkg}, the index computation was extended to other dimensions and to more generic scenarios where our assumption \eqref{no-holo} of no holomorphic vector fields and stable gauge bundles does not hold, such as for example principal $T^2$ fibrations over $K3$ manifolds, see for example \cite{Dasgupta_1999,Goldstein_2004,fu2008theory, Becker:2009df,Melnikov:2014ywa,GarciaFernandez2020} and references therein. These geometries are far from \K Calabi-Yau and it would be interesting to see how the vanishing of the Euler characteristic connects to the physics of these vacua. 

Finally, there has been a lot of recent developments in understanding higher order Yukawa couplings of moduli and matter fields in heterotic compactifications \cite{Buchbinder:2016jqr, McOrist:2016cfl, Ashmore:2018ybe, Anderson:2020ebu, Anderson:2021unr, Anderson:2022kgk, Butbaia:2024tje, Constantin:2024yxh, Ibarra:2024hfm, Gray:2024xun}. Given its topological nature, it is natural to wonder if the vanishing Euler characteristic has something to say about the true nature of the heterotic moduli problem (massless spectrum), once all higher order and non-perturbative corrections have been included.

\subsection*{Acknowledgements}
We would like to thank Xenia de la Ossa, Hannah de Lazari, Mario Garcia-Fernandez, Raul Gonzalez Molina, Jason Lotay, Javier José Murgas Ibarra, Henrique Sa Earp, and Markus Upmeier for interesting conversations. JM is partially supported by a ARC Discovery Grant DP240101409. SP is supported by an NSERC Discovery Grant. JM and ES would like to thank the mathematical research institute MATRIX in Australia where part of this research was performed.


\providecommand{\href}[2]{#2}\begingroup\raggedright\endgroup

\end{document}